\documentclass[prl,aps, amsfonts, 11pt, nofootinbib, notitlepage]{revtex4-1}

\usepackage{graphicx}
\usepackage{subfigure}
\usepackage{hyperref}
\usepackage{epsfig}
\usepackage{amssymb}
\usepackage{amsthm}
\usepackage{amsmath}
\usepackage{amsmath} 
\usepackage{color}

\newcommand{\beq}{\begin{equation}}
\newcommand{\eeq}{\end{equation}}
\newcommand{\bea}{\begin{eqnarray}}
\newcommand{\eea}{\end{eqnarray}}

\newcommand{\nn}{\nonumber}
\newcommand{\singlet}{{^1}\hspace{-.06 cm}{S_0}}
\newcommand{\triplet}{{^3}\hspace{-.06 cm}{S_1}}
\newcommand{\pinot}{\not\hspace{-.08cm}{\pi}}

\newcommand{\benn}{\begin{displaymath}}
\newcommand{\eenn}{\end{displaymath}}

\begin{document}

\begin{flushright}
{NT@UW-12-05}
\end{flushright}

\title{Moving Multi-Channel Systems in a Finite Volume \\
with Application to Proton-Proton Fusion}
\author{Ra\'ul A. Brice\~no\footnote{{\tt briceno@uw.edu}} and Zohreh Davoudi
\footnote{{\tt davoudi@uw.edu}}}
\affiliation{Department of Physics, University of Washington\\
Box 351560, Seattle, WA 98195, USA}

\begin{abstract}
The spectrum of a system with multiple channels composed of two hadrons with nonzero total momentum is determined in a finite cubic volume with periodic boundary conditions using effective field theory methods. The results presented are accurate up to exponentially suppressed corrections in the volume due to the finite range of hadronic interactions. The formalism allows one to determine the phase shifts and mixing parameters of $\pi\pi-KK$ isosinglet coupled channels directly from Lattice Quantum Chromodynamics. We show that the extension to more than two channels is straightforward and present the result for three channels. From the energy quantization condition, the volume dependence of electroweak matrix elements of two-hadron processes is extracted. In the non-relativistic case, we pay close attention to processes that mix the $\singlet-\triplet$ two-nucleon states, e.g. proton-proton fusion ($pp\rightarrow d+e^++\nu_e$), and show how to determine the transition amplitude of such processes directly from lattice QCD. 

 \end{abstract}
\maketitle

\section{Introduction}

Scattering processes in hadronic physics provide useful information
about the properties of particles and their interactions. Some of
these processes are well investigated in experiments with reliable
precision. 
However, there are interesting two-body hadronic processes whose experimental determinations continue to pose challenges. They mainly include two-body
hadronic scatterings near or above the kinematic threshold with the
possibility of the occurrence of resonances. Here we discuss two pertinent cases in Quantum Chromodynamics (QCD), the first of which is the scalar sector, whose nature is still puzzling (see for example Ref.~\cite{coup3} and references therein). While some phenomenological models
suggest the scalar resonances to be tetraquark states (as first proposed by Jaffe \cite{fourquark}), others
propose these to be weakly bound mesonic molecular states. The
most famous of which are the flavorless $a_{0}(980)$ and $f_{0}(980)$, that
are considered to be candidates for a $K\bar{K}$ molecular states
\cite{kkmol, uchipt, kkmol2}. 
In order to shine a light on the nature of these resonances, it would be necessary to perform model-independent multi-channel calculations including the  $\{\pi\pi,\pi\pi\pi\pi, K\bar K, \eta\eta\}$ scattering states directly from the underlying theory of QCD. 
In the baryonic sector, observations of the strong attractive nature of the isosinglet $\bar{K}N$ scattering channel led to the postulation of kaon condensation in dense nuclear matter \cite{Kaplan}. However, extracting $\bar{K}N$ scattering parameters is a rather challenging task, due to the presence of the $\Sigma\pi$ scattering channel and the $\Lambda\left(1405\right)$ resonance below the $\bar{K}N$ threshold, see for example Ref. \cite{LambdaSavage}. Previous chiral perturbation theory calculations have found inconsistency between experimental determination of the $\bar{K}N$ scattering length from scattering data and kaonic hydrogen level shifts \cite{KHydrogen,KHydrogen1, KHydrogen2, KHydrogen3, KHydrogen4}, but as with any low-energy effective field theory (EFT) calculation, there are unaccountable systematic errors associated with the large number of unknown low-energy coefficients (LECs) needed to perform accurate calculations of multi-channel processes.

In addition to these strongly coupled scattering processes, there
are weak processes involving multi-hadron states that require further investigation. For instance, Lattice QCD (LQCD) calculations
have recently shown further evidence of $\Xi^{-}\Xi^{-}$ and $\Lambda\Lambda$
($H$-dibaryon) shallow bound states \cite{nplqcd1,nplqcd2, halqcd}. This
will certainly reignite experimental searches for evidences of these
states. Among the possible weak decays of the $H$-dibaryon include
$H\rightarrow(n\Lambda,n\Sigma^{0},p\Sigma^{-},nn)$ \cite{hdibaryon_decay}. 
In hyper-nuclear physics, there has been much interest in a definitive
determination of the contribution of non-mesonic weak decays ($\Lambda N\rightarrow NN$, $\Lambda NN\rightarrow NNN$) to
the overall decay of hyper-nuclei. In particular, as discussed in Ref. \cite{hyperon} (and references therein), there was a long standing puzzle regarding the theoretical underestimation of the ratio of the decay widths $\Gamma(\Lambda n\rightarrow nn)/\Gamma(\Lambda p\rightarrow np)$ as compared to the experimental value. Certainly a great deal of progress has been made by meson-exchange models in order to close this gap, however a model-independent calculation directly from QCD would give further insight into the mechanism of these decays.  These two cases illustrate processes where it is necessary to evaluate weak matrix elements between multi-hadronic states.

 Currently, LQCD provides the most reliable method for performing calculations of low-energy QCD observables. LQCD calculations are necessarily performed in a Eucledian and finite spacetime volume.  Although the former forbids one to calculate the physical scattering
amplitudes from their Euclidean counterparts away from the kinematic
threshold due to the Maiani-Testa theorem \cite{Maiani}, the latter is proven
to be a useful tool in extracting the physical scattering quantities
from lattice calculations. In his prominent work, L\"uscher showed
how one can obtain the infinite volume scattering phase shifts by
calculating energy levels of interacting two-body systems in the finite
volume \cite{luscher1,luscher2}. The L\"uscher method which was later generalized to the moving frames in Refs. \cite{movingframe, sharpe1, movingframe2}, is only applicable to scattering processes below the inelastic threshold. Therefore it cannot be used near the inelastic
threshold where new channels open up, and a generalized formalism
has to be developed to deal with the coupled (multi)-channel processes.
A direct calculation of the near threshold scattering quantities using
LQCD can lead to the identification of resonances in QCD such
as those discussed above, and provide reliable predictions for their
masses and their decay widths. One such generalization was developed by
Liu $et\; al.$ in the context of quantum mechanical two-body scattering
\cite{coup02, coup2}. There, the authors have been able to deduce the relation between
the infinite volume coupled-channel S-matrix elements and the energy
shifts of the interacting particles in the finite volume by solving
the coupled Schrodinger equation both in infinite volume and on a
torus. The idea is that as long as the exponential volume corrections
are sufficiently small, the polarization effects, as well as other
field theory effects, are negligible. Therefore after replacing the non-relativistic (NR)
 dispersion relations with their relativistic counterparts, the quantum
mechanical result of Liu $et\; al.$ \cite{coup02, coup2} is speculated to be
applicable to the massive field theory. In another approach, Lage
$et\; al.$ considered a two-channel Lippman-Schwinger equation in
a NR effective field theory. They presented the mechanism for obtaining the $\bar{K}N$ scattering length, and studying the nature of the $\Lambda\left(1405\right)$ resonance from LQCD \cite{coup1}. Later on, Bernard $et\; al.$ generalized
this method to the relativistic EFT which would be applicable for coupled
$KK-\pi\pi$ channels \cite{coup3}. Unitarized chiral perturbation theory provides another tool to study a variety of resonances in the coupled-channel scatterings. This method uses the Bethe-Salpeter equation for a coupled-channel system to dynamically generate the resonances in both light-meson sector and meson-baryon sector in the infinite volume, see for example Refs. \cite{uchipt, Kaiser, Locher, OsetII}. When applied in the finite volume, the volume-dependent discrete energy spectrum can be produced, and by fitting the parameters of the chiral potential to the measured energy spectrum on the lattice, the resonances can be located by solving the scattering equations in the infinite volume. This method has been recently used to study the resonances $f_0(600)$, $f_0(980)$ and $a_0(980)$ in Refs. \cite{coup2011, Oset}, $\Lambda(1405)$ in Ref. \cite{MartinezTorres}, $a_1(1260)$ in Ref. \cite{Roca}, $\Lambda_c(2595)$ in Ref. \cite{Xie}, and $D^*_{s 0}(2317)$ in Ref. \cite{MartinezTorresII} in the finite volume. One should note that in contrast with the single-channel scattering system, coupled-channel scattering
requires determination of three independent scattering parameters
which would require in the very least three measurements of the energy
levels in the finite volume. As proposed in Refs. \cite{coup3, coup2011}, one
can impose twisted boundary conditions in the lattice calculation
to be able to increase {the number of measurements by varying the twist angle} and further constrain the scattering parameters. Another tool
to circumvent this problem is the use of asymmetric lattices as is
investigated in Refs. \cite{coup3,coup2011, MartinezTorres}. Alternatively, one can perform calculations with different boost momenta \cite{MartinezTorres,Roca}.

The goal of this paper, as discussed in the following paragraphs, is two-fold. First, in section \ref{mes} we present a model-independent fully relativistic framework for determining the finite volume (FV) coupled-channel spectrum in a moving frame. Secondly, in section  \ref{LL} we show how to extract the matrix element of the current operator between two-hadron states directly from from a FV calculation of the matrix elements, for both relativistic and NR processes. These two are small stepping-stones towards one of the overarching goals of hadronic physics, which is to determine properties of multi-hadron systems directly from the underlying  theory of QCD.

This paper is structured as follows. In the first section we present the result for a scalar field theory model, which illustrates all the features of the problem at hand and allows us to derive the quantization condition (QC) for energy eigenvalues using a diagrammatic expansion. Although, in deriving the QC for this toy model we make a series of approximations, it is shown that this result is in perfect agreement with the exact QC which is obtained from the generalization of the work by Kim $et\; al.$ \cite{sharpe1,sharpe2}.   
In section \ref{N}, we derive the general form of the FV quantization condition for $N$ arbitrarily strongly coupled two-body states. This result has been independently derived and confirmed in a parallel work by Hansen and Sharpe \cite{Hansen:2012tf, Hansen:2012bj}. In this paper, most of the emphasis will be placed on the N=2 case, but the result for the N=3 case will be also explicitly shown.
 
After developing the FV coupled-channel formalism, we extend
our work to be able to determine electroweak matrix elements in the
two-hadron sector. The formalism for extracting the physical transition amplitude for $K\rightarrow \pi\pi$ from the FV
 matrix elements of the weak Hamiltonian in the finite volume, has
been developed by Lellouch and L\"uscher \cite{LL1}. This former quantity
has been shown to be proportional to the latter at the leading order
in the weak coupling, and the proportionality factor (the LL factor), is shown to be related to the derivative of the $\pi\pi$
scattering phase shift. The generalization of the LL factor to the moving frame
is given in Refs. \cite{sharpe1, sharpe2, movingframe2}. Ultimately, it would be desirable to
calculate electroweak matrix elements of states containing an arbitrary
number of hadrons, but for the time being we restrict ourselves to
the two-body sector. Furthermore, we pay close attention to processes involving two nucleons. As is discussed
in Ref. \cite{nnd}, the weak disintegration of deuteron in processes such
as $\nu d\rightarrow\nu pn$ and $\nu_{e}d\rightarrow nne^{+}$ is
of great importance in some neutrino experiments. These processes entail weak
mixing between the $^{1}S_{0}- {^{3}}S_{1}$ two-nucleon channels, and for energies below the pion-production threshold, the dynamics of this system can be described using a pionless EFT (EFT$\left(\pinot\right)$) \cite{pds,pds2,pionless}. 
The result presented gives a relation between the FV weak matrix element and the infinite volume LECs theat parametrize the one-body and two-body weak axial-vector currents. In deriving the result it has been assumed that the FV two-nucleon ground states predominantly described by NR S-wave phase shifts, and we have also used degenerate perturbation theory in the derivation. Both of these are reasonable approximations that will require further investigation to correctly access the size of the corrections.  

\section{Meson-Meson Coupled Channels \label{mes}}

The goal of this section is to present the quantization condition for N coupled-channel system in a moving frame. We present two independent ways to obtain the quantization condition desired. In section \ref{toy}, we present a toy model that illustrates the features of the problem at hand and allows us to determine the scattering matrix using a diagrammatic expansion. In order to derive the quantization condition using this approach, it is convenient to consider the case where the parameter responsible for mixing two channels is small. However, as shown in section \ref{N}, the result presented is in fact exact for an arbitrary mixing parameter. 

In section \ref{N}, we show how to obtain the quantization condition for N arbitrarily strongly coupled two-body channels in a moving frame by solving the Bethe-Salpeter equation. This case is the most relevant if we want to consider systems coupled via QCD interactions, e.g. the $I=0$ system $\pi\pi\rightarrow(K\bar{K},\eta\eta)\rightarrow \pi\pi$.\footnote{It is important to mention that we are not making any claims about the relevance of the four-pion channel in the light-scalar sector of QCD. At this point, it is not evident how to incorporate such states into the calculation, therefore if one would choose to use the formalism presented here to study the light-scalar sector of QCD, there will be an overall systematic error that must be accounted for through other means.}

\subsection{Toy problem: two weakly coupled two-meson channels in the moving frame  \label{toy}}
In this section, we consider a two-meson coupled system with total angular momentum equal to zero.  The two channels will be labeled $I$ and $II$. In general the four mesons can have different masses and quantum numbers, but for the time being we restrict ourselves to the case where two mesons in  each channel are  identical.  Using the ``barred" parameterization \cite{Smatrix}, the time-reversal invariant S-matrix describing this system can be written as
\begin{eqnarray}
\label{smatrix2}
S_2=\begin{pmatrix} 
e^{i2\delta_I}\cos{2\overline{\epsilon}}&ie^{i(\delta_I+\delta_{II})}\sin{2\overline{\epsilon}}\\
ie^{i(\delta_I+\delta_{II})}\sin{2\overline{\epsilon}}&e^{i2\delta_{II}}\cos{2\overline{\epsilon}} \\
\end{pmatrix},
\end{eqnarray}
where $\delta_{I}$ and $\delta_{II}$ are the phase shifts corresponding to the scattering in channels  $I$ and $II$ respectively, and $\bar{\epsilon}$ is a parameter which characterizes the mixing between the channels.  The subscript $2$ on $S$ denotes the number of coupled channels.

At energies below the four-meson threshold, the dynamics of such system can be described by a simple scalar effective field theory (EFT)
\begin{eqnarray}
\mathcal{L}=\sum_{i={I,II}} \phi_{i}^\dag(\partial^2-m_i^2)\phi_{i}-
 \left(\frac{\phi_I\phi_I}{2} \hspace{.3cm}\frac{\phi_{II}\phi_{II}}{2}\right)^\dag\begin{pmatrix} 
c_I&{g}/{2}\\
{g}/{2}&c_{II} \\
\end{pmatrix}
\begin{pmatrix} 
{\phi_{I}\phi_{I}}/{2}\\
{\phi_{II}\phi_{II}}/{2} \\ 
\end{pmatrix}+\cdots,
\end{eqnarray} 
where $\phi_i$  is the  meson annihilation operator for the $i^{th}$ channel, and $c_i$ and g are the LECs of the theory. Ellipsis denotes higher derivative four-meson terms which will be neglected in this section. Our goal is to determine the coupled-channel spectrum in a finite cubic volume with the periodic boundary conditions (PBCs) that falls within the p-regime of LQCD \cite{pregime,pregime2}, defined by $\frac{m_\pi L}{2\pi}\gg 1$ where $m_\pi$ is the pion mass, and $L$ is the spatial extent of the volume. In this regime, FV corrections to the single-particle dressed propagator are exponentially suppressed \cite{luscher1}, as are FV contributions from t- and u-channel scattering diagrams  \cite{luscher1,luscher2,LL1}. The leading order (LO) volume effects in the two-body sector arise from the presence of poles in the s-channel scattering diagrams. These lead to power-law volume corrections to the two-particle spectrum \cite{luscher1,luscher2}. Therefore in the following discussion we will restrict ourselves to the contribution of such diagrams to the scattering matrix $\mathcal{M}$, whose $\mathcal{M}_{ij}$ matrix element corresponds to the scattering amplitude from the $i^{th}$ channel to the $j^{th}$ channel. Certainly, the t (u)-channel diagrams contribute to the renormalization of the theory and therefore to the definition of the LECs, but for energies below the four-particle threshold, their FV effects are exponentially suppressed. It is convenient to redefine the LECs to absorb the contributions from all diagrams with exponentially suppressed FV effects, such as those arising from the t- and u-channel diagrams. 

When considering momentum-independent interactions, the infinite volume loops contributing to the s-channel diagrams are
\begin{eqnarray}
\label{loop1}
G^{\infty}_{i}\equiv\frac{i}{2}\int\frac{d^4k}{(2\pi)^4}\frac{1}{\left[(k-P)^2-m_i^2+i\epsilon\right][k^2-m_i^2+i\epsilon]},
\end{eqnarray}
where $P=(E,\textbf P)$ is the total four-momentum of the system.
\begin{figure}
\begin{centering}
\includegraphics[totalheight=6cm]{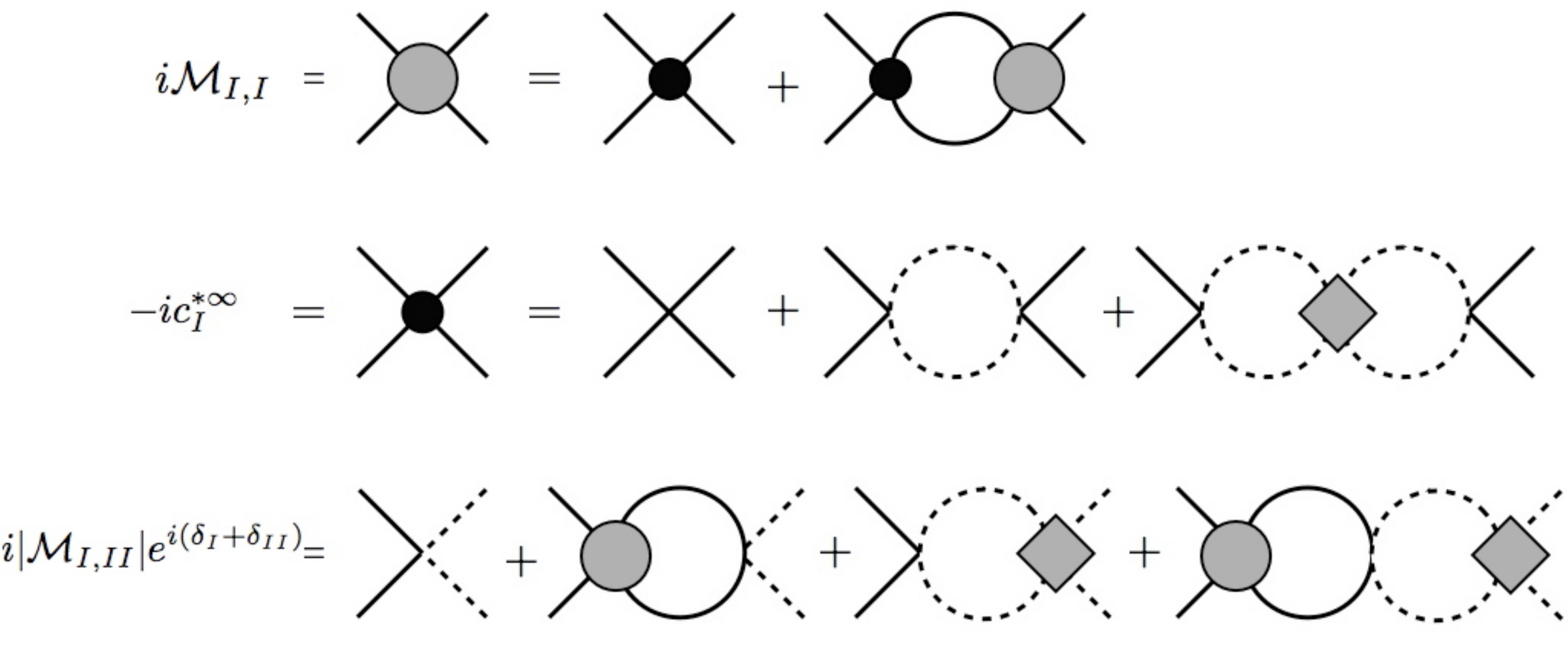}
\par
\caption{Shown are the diagrammatic  representations of the full scattering amplitudes. Solid (dashed) lines represent single particle propagators of the $I$ ($II$) channel. The first two diagrams require the LECs to reproduce the scattering amplitude for the $I$ channel. The grey circle (diamond) is the full scattering amplitude of the $I$ ($II$) channel. The black dot denotes the effective coupling for the $I$ channel, which includes an infinite series of intermediate  $II$ bubble diagrams. There are two other diagrams for the $II$ channel that require the LECs to recover $\mathcal{M}_{II,II}$ and have not been shown. The third diagram ensures that the mixing term $g$ reproduces the off-diagonal scattering matrix elements. }\label{coupledeq}
\end{centering}
\end{figure}
It is straightforward to write down the equations that $\{c_I,c_{II},g\}$ must simultaneously satisfy in order to reproduce the $\mathcal{M}$-matrix elements of the theory, similar to the approach of Ref. \cite{Delta} in examining the role of $\Delta\Delta$ intermediate states in the $\singlet$ (nucleon-nucleon) NN scattering. These equations, which are diagrammatically shown in Fig. (\ref{coupledeq}), can be written as follows
\begin{eqnarray}
\label{reg}
c^{*\infty}_{i} &=&c_{i}+g^2\sum_{j\neq i}\frac{G^{\infty}_{j}}{1-c^{*\infty}_{j}G^{\infty}_{j}}=-\frac{\mathcal{M}_{i,i}}{1-\mathcal{M}_{i,i}G^{\infty}_{i}},
\\ 
g=&-&|\mathcal{M}_{I,II}|e^{i(\delta_I+\delta_{II})} {({1-c^{*\infty}_{I}G^{\infty}_{I}})({1-c^{*\infty}_{II}G^{\infty}_{II}})},
\end{eqnarray}
where $\mathcal{M}_{i,i}$ is the full relativistic S-wave scattering amplitude for the $i^{th}$ channel,  and $\mathcal{M}_{I,II}$ is the amplitude describing the mixing between the two channels. $c^{*\infty}_{i(j)}$ is an effective coupling for the $i (j)$ channel, which includes an infinite series of intermediate $j (i)$ bubble diagrams as shown in Fig. (\ref{coupledeq}). Solving these coupled equations leads to a renormalized theory and determines the scale-dependence of the LECs. 
 
Once this is done, one can study physics in a finite volume. In particular, we are interested in the energy eigenvalues of the meson-meson system placed in a finite volume with the PBCs. The spectrum can be determined by requiring the real part of the inverse of the FV scattering amplitude to vanish.\footnote{One should note that using the notion of FV scattering amplitude is merely for the mathematical convenience. As there is no asymptotic state by which one could define the scattering amplitude in a finite volume, one should in principle look at the pole locations of the two-body correlation function. However, one can easily show that both correlation function and the so-called FV scattering amplitude have the same pole structure, so we use the latter for the sake of simpler representation.} In a periodic volume, the integrals over the spatial momenta appearing in Feynman diagrams are replaced by a sum over discretized three-momenta. In finite volume the integral in Eq. (\ref{loop1}) is replaced by 
\begin{eqnarray}
\label{loop2}
G^{V}_{i}\equiv\frac{i}{2L^3}\sum_{\mathbf{k}}\int\frac{dk^0}{2\pi}\frac{1}{\left[(k-P)^2-m_i^2+i\epsilon\right][k^2-m_i^2+i\epsilon]},
\end{eqnarray}
where the spatial momenta are quantized due to the PBCs $\mathbf k=2\pi \mathbf n/L$ for $\textbf{n}\in Z^3$, while the temporal extent of the Minkowski space remains infinite. This sum suffers from the same UV divergence of Eq. (\ref{loop1}), and the difference of the two,  $\delta{G^{V}_{i}}\equiv{G^{V}_{i}}-{G^{\infty}_{i}}$, is finite.

It is simplest to consider the case where the mixing coupling, $g$ is small, and keep our expressions to leading order in $g$. Using this and the definitions of the LECs in Eq. (\ref{reg}), the FV scattering amplitude of the channel $I$ can be written as

\begin{eqnarray} 
\label{fvscat}
(\mathcal{M}_{I,I})_V\approx
\frac{i\mathcal{M}_{I,I}}{1+\mathcal{M}_{I,I}\delta G^{V}_{I}+\delta G^{V}_{II}\frac{|\mathcal{M}_{I,II}|^2 e^{i2(\delta_I+\delta_{II})}}{\mathcal{M}_{I,I}(1+\mathcal{M}_{II,II}\delta G^{V}_{II})}}.
\end{eqnarray}

Finally, we obtain the quantization condition for the coupled-channel problem at LO in the mixing parameter
\begin{eqnarray}
\label{LOquant}
\mathcal{R}e\left\{|\mathcal{M}_{I,II}|^2 e^{i2(\delta_I+\delta_{II})}
-\left(\mathcal{M}_{I,I}+\frac{1}{\delta G^{V}_{I}}\right)\left(\mathcal{M}_{II,II}+\frac{1}{\delta G^{V}_{II}}\right)\right\}=0,
\end{eqnarray}
which is equivalent to the result of Ref. \cite{Hansen:2012tf, Hansen:2012bj}.\footnote{Note that the FV loop function for channel $i$, $F_{i}$, defined in Eq. (24) of Ref. \cite{Hansen:2012tf, Hansen:2012bj} is equal to $\left(-i\right)$
  times the FV loop function $\delta G_{i}^{V}$
  as defined in Eq. (\ref{def0}) below.} At this point we have refrained from using explicit expressions for the FV integrals and scattering matrix elements; these details will be presented in the following section. It is important to note that despite the simplicity of this toy model, it illustrates all the features of the problem under the investigation, and as will be shown in the next section, the result for the weakly coupled channels presented above is the same to the strongly coupled case when only the S-wave contribution to the two-body scattering in the cubic lattice is taken into account.

\subsection{N strongly coupled two-body channels in a moving frame \label{N}} In the previous section, the following assumptions have been made. First of all, the higher order four-meson terms in the derivative expansion of the effective Lagrangian have been neglected. Secondly, the scattering is restricted to two coupled channels composed of identical particles, and only the S-wave scattering is considered. Most importantly, the mixing term was assumed to be small in deriving the quantization condition. As mentioned earlier, the latter condition is the most relevant when considering systems coupled via QCD interactions. In this section we will simultaneously remove all of this assumptions. 

Most of the details associated with generalizing to a moving frame have been developed by Kim $et\; al.$ \cite{sharpe1,sharpe2}, which will be briefly reviewed here for completeness  (see also Refs. \cite{movingframe, movingframe2} for alternative derivations). A system with total energy and momentum, $E$ and $\mathbf{P}$, in the laboratory frame has a CM energy  $E^*=\sqrt{E^2-\mathbf{P}^2}$. For the $i^{th}$ channel with two mesons each having masses $m_{i,1}$ and $m_{i,2}$, the CM relative momentum is
\begin{eqnarray}
\label{momentum}
q^{*2}_i=\frac{1}{4}\left(E^{*2}-2(m_{i,1}^2+m_{i,2}^2)+\frac{
(m_{i,1}^2-m_{i,2}^2)^2}{E^{*2}}\right),
\end{eqnarray}
which simplifies to $\frac{E^{*2}}{4}-m_{i}^2$ when $m_{i,1}=m_{i,2}=m_{i}$. 

\begin{figure}[t] 
\begin{center}
\subfigure[]{
\label{twopar_FV1}
\includegraphics[scale=0.40]{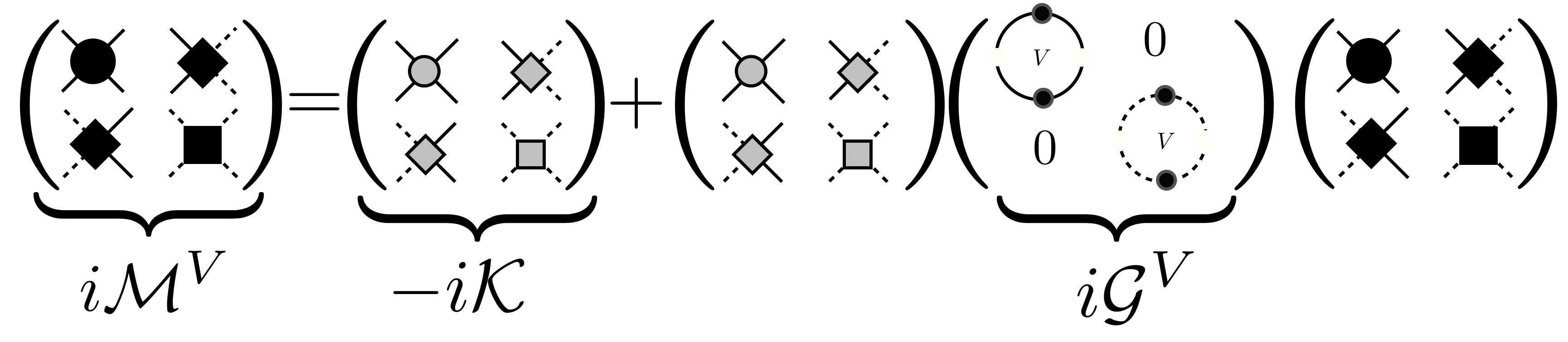}}
\subfigure[]{
\includegraphics[scale=0.23]{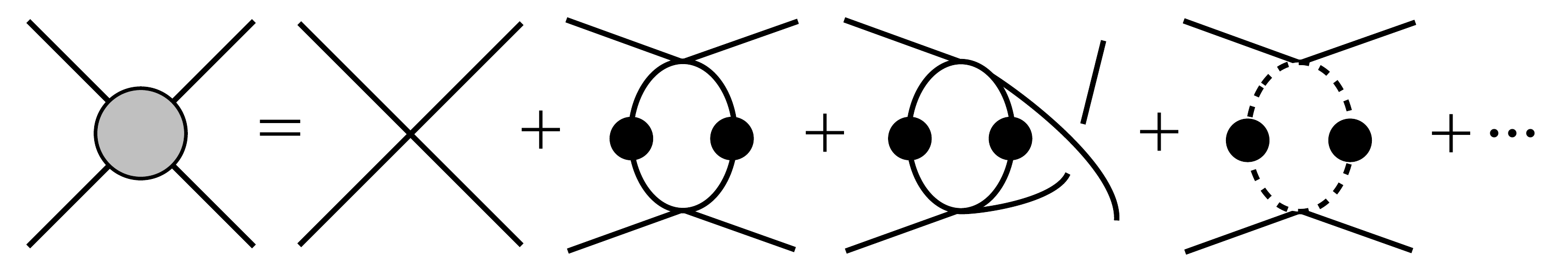}
\label{twopar_FV2}}
\subfigure[]{
\includegraphics[scale=0.23]{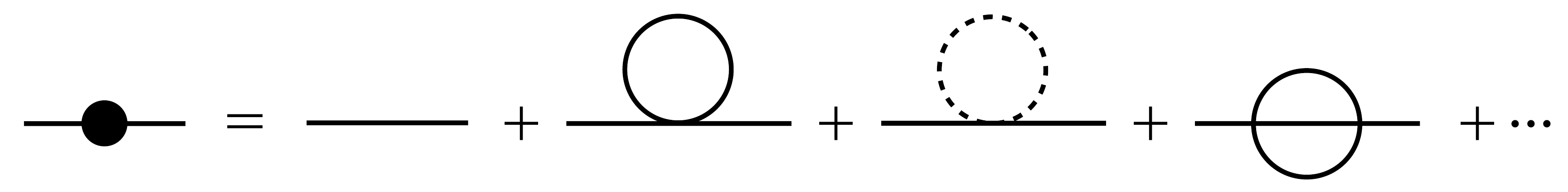}
\label{twopar_FV3}}
\caption{a) The fully-dressed FV two-particle propagator, $\mathcal{M}^V$ can be written in a self-consistent way in terms of the Bethe-Salpeter Kernel, $\mathcal{K}$ and the FV s-channel bubble $\mathcal{G}^V$. 
Note that the only difference between the diagrammatic expansion of the scattering amplitude in the coupled-channel case and the single-channel case is that now the amplitude, kernels and two-particle propagators should be promoted to matrices in the basis of the channels that are kinematically allowed. b) Shown is the $\mathcal{K}_{I,I}$-component of the kernel, which sums all s-channel two-particle irreducible diagrams for channel $I$. c) The fully dressed one-particle propagator is the sum of all one-particle irreducible diagrams and is denoted by a black dot on the propagator lines.}
\label{twoparFV4}
\end{center}
\end{figure}

Because the $S$-matrix for the $l^{th}$ partial wave is a N-dimensional matrix, the scattering amplitude is also necessarily a N-dimensional matrix. In order to have a fully relativistic result that holds for all possible energies below four-particle threshold, the scattering amplitude must include all possible diagrams, i.e. contributions from s-, t- and u-channels as well as self-energy corrections. Fig. \ref{twopar_FV1} depicts the FV analogue of the scattering amplitude, $\mathcal{M}^V$, for the special case of $N=2$ channels. This amplitude is written in a self-consistent way in terms of the Bethe-Salpeter kernel, $\mathcal{K}$, which is the sum of all s-channel two-particle irreducible diagrams, Fig. \ref{twopar_FV2}. For energies below the four-particle threshold, the intermediate particles in the kernel and the self-energy diagrams, Fig. \ref{twopar_FV3}, cannot go on-shell, and therefore these are exponentially close to their infinite-volume counterparts. In fact, only in the s-channel diagrams can all intermediate particles be simultaneously put on-shell. 

Having upgraded the kernel and the two-particle propagators to matrices in the space of open channels, it is straightforward to obtain a non-perturbative quantization condition for the energy levels of the system. It is important to note that the channels only mix by off-diagonal terms in the kernel, which implies that in the absence of interactions a two-pion state continues to propagate as a two-pion state. With this in mind, in the presence of momentum-dependent vertices Eq.~(\ref{loop1}) is replaced by 
\begin{eqnarray}
\label{loop3}
\left[iG^{V}(\mathbf{p}_a,\mathbf{p}_b)\right]_{ab}&\equiv&\frac{n_i}{L^3}\sum_{\mathbf{k}}\int\frac{dk^0}{2\pi}\frac{[\mathcal{K}(\mathbf{p}_a,\mathbf{k})]_{ai}~[\mathcal{K}(\mathbf{k},\mathbf{p}_b)]_{ib}}{[(k-P)^2-m_{i,1}^2+i\epsilon][k^2-m_{i,2}^2+i\epsilon]},
\end{eqnarray}
where the subscripts $a, i, b$ denote the initial, intermediate and final states, respectively, and $n_i$ is ${1}/{2}$ if the particles in the $i^{th}$ loop are identical and 1 otherwise. The sum over all intermediate states, and therefore index $i$ is assumed.
Since the FV corrections arise from the pole structure of the intermediate two-particle propagator, one would expect that the difference between this loop and the its infinite volume counterpart should depend on the on-shell momentum. The on-shell condition fixes the magnitude of the momentum running through the kernels but not its direction. Therefore it is convenient to decompose the product of the kernels into spherical harmonics. These depend not only on the directionality of the intermediate momentum but also on those of the incoming and outgoing momenta. Also, one may represent the N two-body propagators as a diagonal matrix $\mathcal{G}=diag(\mathcal{G}_{1},\mathcal{G}_{2},\cdots, \mathcal{G}_{N})$ as depicted in Fig. \ref{twopar_FV1}. These are infinite-dimensional matrices with with matrix elements \cite{sharpe1,sharpe2}
\begin{eqnarray}
(\delta G^{V}_i)_{l,m;l',m'}\equiv
(G^{V}_i-G^{\infty}_i)_{l,m;l',m'}=
-i\left(\mathcal K \delta \mathcal G^{V}_i \mathcal K\right)_{l,m;l',m'},
\end{eqnarray}
where 
\begin{eqnarray}
\label{def0}
(\delta \mathcal{G}^{V}_{i})_{l_1,m_1;l_2,m_2}&=&i	\frac{q^*_in_i}{8\pi E^*}\left(\delta_{l_1,l_2}\delta_{m_1,m_2}+i\frac{4\pi}{q_i^*}\sum_{l,m}\frac{\sqrt{4\pi}}{q_i^{*l}}c^{\textbf{P}}_{lm}(q_i^{*2})\int d\Omega^*Y^*_{l_1m_1}Y^*_{lm}Y_{l_2m_2}	\right),\nn\\
\end{eqnarray}
and the function $c^{\textbf{P}}_{lm}$ is defined as\footnote{Note that our definition of the $c_{lm}^{\textbf{P}}$ function differs that of Ref. \cite{sharpe1} by an overall sign.}
\begin{eqnarray}
\label{clm}
c^{\textbf{P}}_{lm}(x)=\frac{1}{\gamma}\left[\frac{1}{ L^3}\sum_{\textbf{k}}-\mathcal{P}\int\frac{d^3\mathbf{k}}{(2\pi)^3}\right]\frac{\sqrt{4\pi}Y_{lm}(\hat{k}^*)~k^{*l}}{{k}^{*2}-x} \ .
\end{eqnarray}
$\mathcal{P}$ in this relation denotes the principal value of the integral, and $\mathbf{k}^*={\gamma}^{-1}(\mathbf k_{||}-\alpha \mathbf P)+\mathbf k_{\perp}$, where $\mathbf k_{||}$ ($\mathbf k_{\perp}$) denotes the component of momentum vector $\mathbf{k}$ that is parallel (perpendicular) to the boost vector $\mathbf{P}$, $\alpha=\frac{1}{2}\left[1+\frac{m_1^2-m_2^2}{E^{*2}}\right]$, $E^*$ is the CM energy of the system, $E^*=\sqrt{q^*+m_1^2}+\sqrt{q^*+m_2^2}$, and the relativistic $\gamma$ factor is defined by $\gamma=E/E^*$~\cite{Davoudi, Fu:2011xz, Sasa}. This reduces to the NR value of $\alpha=\frac{m_1}{m_1+m_2}$ as is presented in Ref.~\cite{Bour:2011ef}. Note that this result is equivalent to the NR limit of the result obtained in Refs. \cite{movingframe, sharpe1,movingframe2} for the boosted systems of particles with identical masses.\footnote{The kinematic function $c^{\textbf{P}}_{lm}(q_i^{*2})$ can also be written in terms of the three-dimensional Zeta function, $\mathcal{Z}^d_{lm}$,
 \begin{eqnarray}
 \nonumber
 	c^{\textbf{P}}_{lm}(q^{*2})=\frac{\sqrt{4\pi}}{\gamma L^3}\left(\frac{2\pi}{L}\right)^{l-2}\mathcal{Z}^d_{lm}[1;(q^*L/2\pi)^2],\hspace{1cm} 
\mathcal{Z}^d_{lm}[s;x^2]=\sum_{\mathbf r \in P_d}\frac{Y_{l,m}(\mathbf{r})}{(r^2-x^2)^s},
\end{eqnarray}
where the sum is performed over $P_d=\left\{\mathbf{r}\in \textbf{R}^3\hspace{.1cm} | \hspace{.1cm}\mathbf{r}={\gamma}^{-1}(\mathbf m_{||}-\alpha \mathbf d)+\mathbf m_{\perp} \text{,}m\in \textbf{Z}\right\}$, $\mathbf d$ is the normalized boost vector $\mathbf d=\mathbf{P}L/2\pi$,  and $\alpha$ is defined above~\cite{Davoudi, Fu:2011xz, Sasa}.}

The generalization of the quantization conditions for N channels that are coupled via an arbitrarily strong interaction is now straightforward given such upgrading of the kernel, $\mathcal K$, to not just be a matrix over angular momentum but also over N channels as discussed before. The kernel is assured to reproduce the infinite volume scattering matrix ($\mathcal{M}$) by solving the following matrix equation
\begin{eqnarray}
i\mathcal{M}&=&-i\mathcal{K}
-i\mathcal{K}\mathcal{G^{\infty}} \mathcal{K}
-i\mathcal{K}\mathcal{G^{\infty}} \mathcal{K}\mathcal{G^{\infty}}\mathcal{K}+\cdots=-i\mathcal{K}\frac{1}{1-\mathcal{G^{\infty}}\mathcal{K}}
\Rightarrow
\mathcal{K}=-\mathcal{M}\frac{1}{1-\mathcal{G^{\infty}}\mathcal{M}}.\end{eqnarray}
With this definition of the kernel, one can proceed to evaluate poles of the  N-channels FV scattering matrix by replacing the infinite volume loops $\mathcal{G}^\infty$ with their FV $\mathcal{G}^V$ counterparts, 
\begin{eqnarray}
-i\mathcal{M}^{V}&=&-i\mathcal{K}-i\mathcal{K}\mathcal{G}^{V} \mathcal{K}-i\mathcal{G}^{V}\mathcal{K}\mathcal{G}^{V}\mathcal{K}+\cdots=-i\mathcal{K}\frac{1}{1-\mathcal{G}^{V}\mathcal{K}}
\\
& = & -i\frac{1}{1-\mathcal{M}\mathcal{G}^{\infty}}\mathcal{M}\frac{1}{1+\delta \mathcal{G}^{V}\mathcal{M}}({1-\mathcal{M}\mathcal{G}^{\infty}}). 
\end{eqnarray} 
Finally arriving at the quantization condition
\begin{eqnarray}
\label{det0}
\mathcal{R}e\left\{\det(\mathcal{M}^{-1}+\delta \mathcal{G}^{V})\right\}=\mathcal{R}e\left\{{\rm{det}}_{\rm{oc}}\left[\rm{det}_{\rm{pw}}\left[\mathcal{M}^{-1}+\delta \mathcal{G}^{V}\right]\right]\right\}=0 ,
\end{eqnarray} 
where the determinant $\rm{det}_{\rm{oc}}$ is over the N open channels and the determinant $\rm{det}_{\rm{pw}}$ is over the partial waves, and both $\mathcal{M}$ and $\delta \mathcal{G}^V$ functions are evaluated on the on-shell value of the momenta. We have taken the real part of the determinant in Eq. (\ref{det0}), but as it will be shown shortly, this determinant condition gives rise to only one single real condition for both single channel and two coupled-channel cases with $l_{max}=0$, so we omit the notion of the real part in the QC from now on. For a general proof of the reality of quantization condition with any number of coupled channels see Refs. \cite{Hansen:2012tf, Hansen:2012bj}.

For N=1, one reproduces the result first obtained by Rummukainen and Gottlieb \cite{movingframe} and later confirmed by Kim $et\; al.$ \cite{sharpe1} and Christ \textit{et al.} \cite{movingframe2} for the case of single-channel moving frame two-particle systems as follows. First note that it is convenient to evaluate the determinant using the spherical harmonic basis of $\delta \mathcal{G}^{V}$, Eq. (\ref{def0}), and the on-shell scattering amplitude $\mathcal{M}_i$ \cite{sharpe1}
\begin{eqnarray}
\label{def1}
(\mathcal{M}_{i})_{l_1,m_1;l_2,m_2}&=&\delta_{l_1,l_2}\delta_{m_1,m_2}\frac{8\pi E^*}{n_iq^*_i}\frac{e^{2i\delta^{(l)}_i(q^*_i)}-1}{2i}.
\end{eqnarray}
If  the two equal-mass meson interpolating operator is in the $A_1^+$  irreducible representation of the cubic group, the energy eigenstates of the system have overlap with the $l=0,4,6,\ldots$ angular momentum states at zero total momentum, making the truncation at $l_{max}=0$ a rather reasonable approximation in the low-energy limit. When $\textbf{P}\neq 0$, the symmetry group is reduced, and at low energies the $l=0$ will mix with the $l=2$ partial wave as well as with higher partial waves \cite{movingframe}. For two mesons with different masses, the symmetry group is even further reduced in the boosted frame, making the mixing to occur between $l=0$ and $l=1$ states as well as with higher angular momentum states \cite{Fu:2011xz}. An easy way to see the latter is to note that in contrast with the case of degenerate masses, the kinematic function $c_{lm}^{\textbf{P}}$ as defined in Eq. (\ref{clm}) is non-vanishing for odd $l$ when the masses are different. As a result even and odd angular momenta can mix in the quantization condition. This however does not indicate that the spectrum of the system is not invariant under parity. As long as all  interactions between the particles are parity conserving, the spectrum of the system and its parity transformed counterpart are the same. One should note that the determinant condition, Eq. (\ref{det0}), guarantees this invariance: any  mechanism, for example, which takes an S-wave scattering state to an intermediate P-wave two-body state, would take it back to the final S-wave scattering state, and the system ends up in the same parity state.\footnote{Note that under parity $\mathcal{Z}^d_{lm}\rightarrow\left(-1\right)^{l}\mathcal{Z}^d_{lm}$. Note  also that under the interchange of particles $\mathcal{Z}^d_{lm}\rightarrow\left(-1\right)^{l}\mathcal{Z}^d_{lm}$, so that for degenerate masses the $c_{lm}^{P}$ functions vanish for odd $l$. This is expected since the parity transformation in the CM frame is equivalent to the interchange of particles. However, as is explained above for the case of parity transformation, despite the fact that $\delta \mathcal{G}^{V}$ is not symmetric with respect to the particle masses, the quantization condition is invariant under the interchange of the particles.}

Nevertheless, let us assume that the contributions from higher partial waves to the  scatterings are negligible, so that one can truncate the determinant over the angular momentum at $l_{max}=0$. Then the familiar quantization condition for the S-wave scattering,
\begin{eqnarray}
q_i^*\cot({\delta_i^0})=4\pi c_{00}^{P}(q^{*2}_i),
\end{eqnarray}
is recovered. It is convenient to introduce a pseudo-phase defined by  
\begin{eqnarray}
 \label{pseudophase}
{q^*_i}\cot({\phi^P_i})\equiv -4\pi{ c_{00}^P(q^{*2}_i)}
 \end{eqnarray}
to rewrite the quantization condition as
\begin{eqnarray}
\label{quateq}
\cot({\delta_i})=-\cot({\phi^P_i})\Rightarrow \delta_i+\phi^P_i=m\pi,
\end{eqnarray}
where $m$ is an integer. In this form, the quantization condition is manifestly real.

 For the N=2 case, the expression for the scattering amplitude in Eq. (\ref{def1}) is modified, as it now depends on the mixing angle $\bar{\epsilon}$, and the scattering matrix is no longer diagonal, while still symmetric. By labeling the off-diagonal terms as $\mathcal{M}_{I,II}$, and using the definition of the S-matrix  for the coupled-channel system, Eq. (\ref{smatrix2}), the scattering matrix elements can be written as
\begin{eqnarray}
\label{def}
(\mathcal{M}_{i,i})_{l_1,m_1;l_2,m_2}&=&\delta_{l_1,l_2}\delta_{m_1,m_2}\frac{8\pi E^*}{n_iq^*_i}\frac{\cos(2\bar{\epsilon})e^{2i\delta^{(l_1)}_i(q^*_i)}-1}{2i},\\ 
(\mathcal{M}_{I,II})_{l_1,m_1;l_2,m_2}&=&\delta_{l_1,l_2}\delta_{m_1,m_2}\frac{8\pi E^*}{\sqrt{n_In_{II}q^*_Iq^*_{II}}}\sin(2\bar{\epsilon})\frac{e^{i(\delta^{(l_1)}_I(q^*_I)+\delta^{(l_1)}_{II}(q^*_{II}))}}{2},
\end{eqnarray}
where the usual relativistic normalization of the states is used in evaluating the S-matrix elements. From Eq. (\ref{det0}) one obtains
\begin{eqnarray}
\label{allorders}
\det\begin{pmatrix} 
1+\delta \mathcal{G}^{V}_I\mathcal{M}_{I,I}&\delta \mathcal{G}^{V}_I\mathcal{M}_{I,II}\\
\delta \mathcal{G}^{V}_{II}\mathcal{M} _{I,II}&1+\delta \mathcal{G}^{V}_{II}\mathcal{M}_{II,II}\\
\end{pmatrix}=0,
\end{eqnarray} 
where the determinant is not only over the number of channels but also over angular momentum which is left implicit. In deriving this result we have made no assumption about the relative size of the scattering matrix elements, but when $l_{max}=0$, we  recover the LO result in Eq. (\ref{LOquant}). For $l_{max}=0$ one can use the pseudo-phase definition in Eq. (\ref{pseudophase}) to rewrite the quantization condition in a manifestly real form,
 \begin{eqnarray}
\label{allorders2}
\cos{2\bar{\epsilon}}\cos{\left(\phi^P_1+\delta_1-\phi^P_2-\delta_2\right)}=\cos{\left(\phi^P_1+\delta_1+\phi^P_2+\delta_2\right)},
 \end{eqnarray} 
which is equivalent to the result given in Refs. \cite{coup02, coup2} in the CM frame.\footnote{The agreement between Eq. (\ref{allorders2}) and Eq. (37) of Ref. \cite{coup02} can be achieved by noting that the pseudo-phase  $\phi^P_{i}$ as defined in Eq. (\ref{pseudophase}) is equivalent to the negative $\Delta_{i}$ as defined in Eq. (36) of Ref. \cite{coup02}. On the other hand, the mixing parameter $\overline{\epsilon}$ as defined in Eq. (\ref{smatrix2}) is related to the mixing parameter $\eta_{0}$ defined in Eq. (14) of Ref. \cite{coup02} through $\eta_{0}=\cos2\bar{\epsilon}$.} It is easy to see that in the $\bar \epsilon\rightarrow 0$ limit, one recovers the decoupled quantization conditions for both channels $I$ and $II$, Eq. (\ref{quateq}).

The extension to a larger number of coupled channels is straightforward. As an example, we consider the N=3 case. Unitarity as well as time-reversal invariance allow us to parametrize the S-matrix using three phases shifts $\{\delta_I, \delta_{II}, \delta_{II}\}$ and three mixing angles  $\{\bar\epsilon_1, \bar\epsilon_2, \bar\epsilon_3\}$ 
\begin{eqnarray}
\label{smatrix3}
S_3=\begin{pmatrix} 
e^{i2\delta_I}c_1
&ie^{i(\delta_I+\delta_{II})}s_1c_3
&ie^{i(\delta_I+\delta_{III})}s_1s_3\\
ie^{i(\delta_I+\delta_{II})}s_1c_2&
e^{i2\delta_{II}}\left(c_1c_2c_3-s_2s_3\right)
&
ie^{i(\delta_I+\delta_{III})}
\left(c_1c_2s_3+s_2c_3\right)\\
ie^{i(\delta_I+\delta_{III})}s_1s_2&
ie^{i(\delta_{II}+\delta_{III})}\left(c_1s_2c_3+c_2s_3\right)&
ie^{i2\delta_{III}}\left(c_1s_2s_3-c_2c_3\right)\\
\end{pmatrix},
\end{eqnarray}
where $c_i=\cos(2\bar{\epsilon}_i)$, $s_i=\sin(2\bar{\epsilon}_i)$. Note that in the limit ${\epsilon}_2={\epsilon}_3=0$ the third channel decouples, and one obtains a block diagonal matrix composed of $S_2$ corresponding to the $I-II$ coupled channel, as well as a single element corresponding to the scattering in the uncoupled channel $III$. The spectrum of three-coupled channel is obtained from
\begin{eqnarray}
\label{det3}
\det\begin{pmatrix} 
1+\delta \mathcal{G}^{V}_I\mathcal{M}_{I,I}&
\delta \mathcal{G}^{V}_I\mathcal{M}_{I,II}
&\delta \mathcal{G}^{V}_I\mathcal{M}_{I,III}
\\
\delta \mathcal{G}^{V}_{II}\mathcal{M}_{II,I}&1+\delta \mathcal{G}^{V}_{II}\mathcal{M}_{II,II}&
\delta \mathcal{G}^{V}_{II}\mathcal{M}_{II,III}
\\
\delta \mathcal{G}^{V}_{III}\mathcal{M}_{III,I}&
\delta \mathcal{G}^{V}_{III}\mathcal{M}_{III,II}
&1+\delta \mathcal{G}^{V}_{III}\mathcal{M}_{III,III}
\end{pmatrix}=0,
\end{eqnarray} 
where the scattering matrix elements can be  determined from Eq. (\ref{smatrix3}) using the relationship between the scattering amplitudes and the S-matrix elements,
\begin{eqnarray}
(\mathcal{M}_{i,j})_{l_1,m_1;l_2,m_2}=\delta_{l_1,l_2}\delta_{m_1,m_2}\frac{8\pi E^*}{\sqrt{n_in_{j}q^*_iq^*_{j}}}\frac{(S^{(l_1)}_3)_{i,j}-\delta_{i,j}}{2i}.
\end{eqnarray}

\section{Two-Body electroweak Matrix Elements in a finite volume \label{LL}} 

As discussed in the introduction, electroweak processes in the two-hadron sector of QCD encompass a variety of interesting processes, so it is desirable to calculate the electroweak matrix elements directly from LQCD. One of the very first attempts to develop a formalism for such processes from a FV Euclidean calculation is due to Lellouch and L\"uscher. In  their seminal work \cite{LL1}, they restricted the analysis to $K\rightarrow \pi\pi$ decay in the kaon's rest frame, and showed that the absolute value of the transition matrix element in an Euclidian FV is proportional to the physical transition matrix element. This proportionality factor is known as the LL-factor. This formalism was then generalized to moving frames in Refs. \cite{sharpe1, movingframe2}. Here we present the generalization of Lellouch and L\"uscher formalism to processes where the initial and final states are composed of two-hadron S-wave states. In the relativistic case, the coupled-channel result, Eq. (\ref{LOquant}), is used to derived the $2\rightarrow 2$ LL-factor for boosted systems, while the contributions from one-body currents to the processes are neglected. In section \ref{pionless} we discuss a particular case, namely $pp\rightarrow d+e^++\nu_e$, where the one-body current is in fact the dominant contribution to the weak transition amplitude. Using EFT($\pinot$), the relationship between the FV matrix element and the pertinent LECs of the weak interaction is found. It is shown that in the presence of a one-body current, the FV and infinite volume weak matrix elements are no longer proportional to one another.  Nevertheless, the infinite volume result can still be determined from the FV calculation of matrix elements upon determining the LECs of the EFT describing the process as become evident shortly.
\subsection{Relativistic 2-Body LL-Factor \label{relLL}}
In order to derive the relativistic two-body LL-factor, one first notes that in the absence of the weak interaction, the two states decouple, and as a result the $S$-matrix becomes diagonal. As is pointed out by Lellouch and L\"uscher, there is a simple trick to obtain the desired relation between infinite volume and FV matrix elements by assuming the initial and final states to be nearly degenerate with energy $E^*_0$ (each satisfying Eq. (\ref{quateq})) when there is no weak interaction. Once the perturbative weak interaction is turned on, the degeneracy is lifted, and the energy eigenvalues are
\begin{eqnarray}
\label{deg}
E^*_{\pm}=E^*_0\pm V|\mathcal{M}^V_{I,II}|\equiv E^*_0\pm\Delta E^*, 
\end{eqnarray}
where $\mathcal{M}^V_{I,II}$ is the FV matrix element of the weak Hamiltonian density. 
Consequently, the CM momenta and the scattering phase shifts acquire perturbative corrections of the form
\begin{eqnarray}
\label{pseudomom}
\Delta q^*_i=\frac{1}{4q_i^*}\left(E^*_0-\frac{(m_{i,1}^2-m_{i,2}^2)^2}{ E_0^{*3}}\right)V|\mathcal{M}^V_{I,II}|\equiv \Delta \tilde{q}^*_i \hspace{.1cm}V|\mathcal{M}^V_{I,II}|,
\end{eqnarray}
and
\begin{eqnarray}
\label{pseudomom2}
\Delta \delta_i({q}^*_i )=\delta'_i({q}^*_i ) \Delta \tilde{q}^*_i \hspace{.1cm}V|\mathcal{M}^V_{I,II}|,
\end{eqnarray}
where $\delta'_i({q}^*_i )$ denotes the derivative of the phase shift with respect to the momentum evaluated at the free CM momentum, and $V=L^3$. The perturbed energy necessarily satisfies the quantization condition, Eq. (\ref{allorders}). The generalized LL-factor for $2\rightarrow 2$ scattering is then obtained by Taylor expanding Eq. (\ref{allorders}) to leading order in the weak matrix element about the free energy solution, 
\begin{eqnarray}
|\mathcal{M}^\infty_{I,II}|^2 
&=&V^2 \left\{\Delta \tilde{q}^*_I \Delta \tilde{q}^*_{II}\left(\frac{8 \pi E^*_0}{n_Iq^*_I}\right)\left(\frac{8\pi E^*_{0}}{n_{II}q^*_{II}}\right)
\left({\phi^{P}_{I}}'(q^*_{I})+\delta_{I}'(q^*_{I})\right)\left({\phi^{P}_{II}}'(q^*_{II})+\delta_{II}'(q^*_{II})\right)\right\}|\mathcal{M}^V_{I,II}|^2,\nn\\\label{mesonLL}
\end{eqnarray}
where ${\phi^{P}_{i}}'({q}^*_i )$ denotes the derivative of the pseudo-phase  with respect to the momentum evaluated at the free CM momentum.

Note that we arrived at the generalization of the LL factor for two-body matrix elements using the degeneracy of states argument.  
Lin $et\; al.$ \cite{LL2} showed that  the LL-factor for $K\rightarrow \pi\pi$ can also be derived using the density of states in the large volume limit, and this argument was then generalized by Kim $et\; al.$ \cite{sharpe1} to boosted systems. Here it will be shown that the result in Eq. (\ref{mesonLL}) is also consistent with the derivation based on the Kim $et\; al.$'s work. Let $\sigma_i\left(\mathbf{x},t\right)$ be the two-particle annihilation operator for the $i^{th}$ channel. Then the two particle correlation function in FV can be written as
\begin{eqnarray}
\label{fincor}
C_{\mathbf{P},i}^{V}\left(t\right)&\equiv&\int d^3x\hspace{.1cm}e^{i\mathbf{P}\cdot \mathbf{x}}\left\langle 0\right|\sigma_i\left(\mathbf{x},t\right)\sigma^\dag_i\left(\mathbf{0},0\right)\left|0\right\rangle_V= V\sum_m   e^{-E_mt}\left|\left\langle 0\right|\sigma\left(\mathbf{0},0\right)\left|i;\mathbf{P},m\right\rangle_V \right|^{2}\nn\\
&\stackrel{L\rightarrow \infty }\longrightarrow&
V \int dE\rho_{V,i}(E)e^{-Et}\left|\left\langle 0\right|\sigma\left(\mathbf{0},0\right)\left|i;\mathbf{P},E\right\rangle_V \right|^{2},
\end{eqnarray}
where a complete set of states is being inserted in the first equality. In the second equality, we have introduced the density of states for the $i^{th}$ channel, $\rho_{V,i}(E)$, which is defined as $\rho_{V,i}(E)=dm_i/dE$. Using Eqs. (\ref{quateq}), (\ref{pseudomom}) the density of states can be written as $\rho_{V,i}(E^*)=\left({\phi^{P}_{i}}'(q^*_{i})+\delta_{i}'(q^*_{i})\right)\Delta\tilde{q}^*_{i}/\pi$. In the infinite volume the two-particle correlation function is \cite{LL2}
\begin{equation}
\label{inftycor}
C_{\mathbf{P},i}^{\infty}\left(t\right)=\frac{n_i}{8\pi^{2}}\int dE\frac{q^{*}_i}{E^{*}}e^{-Et}\left|\left\langle 0\right|\sigma\left(\mathbf{0},0\right)\left|i;\mathbf{P},E\right\rangle_\infty \right|^{2}, 
\end{equation}
where the factor of $n_i$ has been introduced to account for the double counting of the phase space when the particles are identical. It is straightforward to show that this relation still holds when the two particles have different masses. From Eqs. (\ref{fincor}), (\ref{inftycor}) the relationship between the states of infinite and asymptotically large (yet finite) volume can be deduced, 
\begin{equation}
\label{FVstates}
\left|i;\mathbf{P},E\right\rangle_\infty\Leftrightarrow
2\pi\sqrt{\frac{2V\rho_{V,i}E_0^*}{n_iq_i^*}}
\left|i;\mathbf{P},E\right\rangle_V.
\end{equation}
This relation therefore recovers the LL-factor as given in Eq. (\ref{mesonLL}). It also demonstrates that the LL-factor accounts for different normalizations of the states in the finite volume and infinite volume in the presence of interactions. 

It is important to understand that this derivation strongly relies on the assumptions that there are no one-body currents that mix the initial and final states. In reality most systems are sensitive to one-body currents. In order to find a relation between the FV and infinite volume electroweak matrix elements it is necessary to understand the contributions from the one-body currents. In the following section we consider one pertinent problem and show that in fact one-body currents introduce additional FV contributions. 


\subsection{Proton-proton fusion in EFT($\pinot$) \label{pionless}}
In this section we discuss the weak interaction in the two-nucleon sector. This sector has been previously studied by Detmold and Savage \cite{nnd} in the finite volume. They considered a novel idea of studying electroweak matrix elements using a background field. Since evaluating matrix elements of electroweak currents between NN states, e.g. $\left\langle d\right|\left.A^{\mu}\right.\left|np\right\rangle 
 $,  is naively one or two orders of magnitude more difficult than performing NN four-point functions, they present a procedure for extracting the relevant LECs of the pionless EFT, EFT($\pinot$) \cite{pionless}, by calculating four-point functions of nucleons in a finite volume in the presence of a background electroweak field.  This would be a project worth pursuing with great benefits, namely a five-point function is replaced by a four-point function, thereby dramatically reducing the number of propagator contractions. For isovector quantities, this procedure comes at a small cost, since for perturbatively small background fields, the QCD generated gauge links get modified by a multiplicative factor that couples the valence quarks to the external field, $U^{QCD}_{\mu}(x)\rightarrow U^{QCD}_{\mu}(x)U_{\mu}^{ext}(x)$. On the other hand, for isoscalar  quantities this approach would require the generation of gauge configurations in the presence of the background field. For both isovector and isoscalar quantities, one would need to perform calculations at a range of background field strengths in order to precisely discern the contribution of the coupling between the background field and the baryonic currents to the NN spectrum. Additionally, the nature of this background field will differ depending on the physics one is interested in.  
Alternatively, one can always evaluate matrix elements of electroweak currents with gauge configurations that solely depend on the QCD action, which is the case considered here. With the improvement in the computational resources available for LQCD calculations, the studies of nucleonic matrix elements will become feasible shortly, and therefore their connection to the physical matrix elements should be properly addressed.

The goal is to explore FV corrections of weak matrix elements in the two-nucleon sector. In particular, we will consider the proton-proton fusion process, $(pp\rightarrow d+e^++\nu_e)$, which mixes the $\singlet-\triplet$ channels. In order to do this calculation the mechanism of EFT($\pinot$) \cite{pds, pds2, pionless} will be used. The methodology is similar to the one used for the toy problem considered in the first section, except the fields are now non-relativistic and carry isospin and spin indices. The presence of a weak interaction, leads to a contribution to the Lagrangian that couples the axial-vector current $A^{\mu=3}=\frac{1}{2}\left(\bar{u}\gamma^3\gamma^5u-\bar{d}\gamma^3\gamma^5d\right)$ to an external weak current. In terms of the low-energy degrees of freedom, the axial current will receive one-body and two-body contributions. At energies well below the pion-production threshold, the  EFT($\pinot$) Lagrangian density including weak interactions can be written as \cite{pds,pds2,pionless,Butler:1999sv, Butler:2000zp, Butler:2001jj}
\begin{eqnarray}
\mathcal{L}&=&N^\dag\left(i\partial_0+\frac{\nabla^2}{2M}-\frac{W_3g_A}{2}\sigma^3\tau^3\right)N
-C^{{\left(\singlet\right)}}_0\left(N^TP_1^aN\right)^\dag \left(N^TP_1^aN\right)\nn\\
&&-C^{{\left(\triplet\right)}}_0\left(N^TP_3^jN\right)^\dag \left(N^TP_3^jN\right)
-{L_{1,A}}W_3\left[\left(N^TP_1^3N\right)^\dag \left(N^TP_3^3N\right)+h.c.\right]+\cdots,
\label{pionlag}
\end{eqnarray}
where $N$ is the nucleon annihilation operator with bare mass $M$,  $\{C^{{\left(\singlet\right)}}_0, C^{{\left(\triplet\right)}}_0, g_A,{L_{1,A}}\}$ are the LECs of the theory, $g_A = 1.26$  is the nucleon axial coupling constant, $W_3$ is the external weak current, and $\{P_1^a, P_3^j\}$ are the standard $\{\singlet,\triplet\}$-projection operators,
\begin{eqnarray}
\label{proj}
P_1^a=\frac{1}{\sqrt{8}}\tau_2\tau^a\sigma_2,\hspace{1cm}
P_3^j=\frac{1}{\sqrt{8}}\tau_2\sigma_2\sigma^j,
\end{eqnarray}
where $\tau$ ($\sigma$) are the Pauli matrices which act in isospin (spin) space. In Eq. (\ref{pionlag}) the ellipsis denotes an infinite tower of higher order operators. The $\mathcal{O}(p^{2n})$-operator for the $\{\singlet, \triplet\}$ state will have  corresponding LECs $\{C^{{\left(\singlet\right)}}_{2n}, C^{{\left(\triplet\right)}}_{2n}\}$, which are included in this calculation. In this section we only consider NN systems in the S-wave channel. As discussed in section~\ref{N} this introduces a systematic error in FV systems that is negligible near the kinematic threshold. 

At leading order, a weak transition between the isosinglet and isotriplet two-nucleon channels is described by an insertion of the single body current (which is proportional to $g_A$) and the bubble chain of the $C_0^{(\triplet)}$ and $C_0^{(\singlet)}$ contact interactions on the corresponding nucleonics legs as discussed in Ref.~\cite{Kong:1999tw}. At next to leading order (NLO), the hadronic matrix element of  $pp\rightarrow d+e^++\nu_e$ receives contributions from one insertion of the the $C_2p^2$ operator along with one insertion of the single-body operator proportional to $g_A$ \cite{Kong:1999mp}. At the same order, a single insertion of the two-body current that is proportional to ${L_{1,A}}$ also contributes to the transition amplitude~\cite{Butler:2001jj}. In both of these contributions the dressing of the NN states with the corresponding bubble chain of the LO contact interactions must be assumed. As is discussed in Ref. \cite{Butler:2001jj}, the two-body contribution is estimated to give rise to a few-percent correction to the hadronic matrix element, and its corresponding LEC, $L_{1A}$, is known to contribute to the elastic and inelastic neutrino-deuteron scattering cross sections as well \cite{Butler:1999sv, Butler:2001jj}. Of course, the electromagnetism plays a crucial role in the initial state interactions in the pp-fusion process, but as is shown in Refs. \cite{Kong:1999tw, Kong:1999mp, Butler:2001jj}, the ladder QED diagrams can be summed up to all orders nonperturbatively. Since LQCD calculations of the matrix elements of the axial-vector current involving two-nucleons would allow for a determination of ${L_{1,A}}$, one will achieve tighter theoretical constraints on the cross section of these processes. Furthermore, having obtained the one-body and two-body LECs of the weak sector will allow for the determination of the few-body weak observables.


In the absence of weak interactions, the on-shell scattering amplitude for both channels can be determined exactly in terms of their corresponding LECs by performing a geometric series over all the bubble diagrams \cite{pionless}
\begin{eqnarray}
\mathcal{M}^0=-\frac{\sum_{n=0}^{\infty} C_{2n}q^{*2n}}{1-G^\infty_0\sum_{n=0}^{\infty} C_{2n}q^{*2n}},
\end{eqnarray}
where the on-shell relative momentum in the CM frame is related to the total NR energy and momentum of the two-nucleon system via, $q^*=\sqrt{ME-\frac{1}{4}{\mathbf{P}^2}}$, and $G^\infty_0$ denotes the loop integral
\begin{eqnarray}
\label{Iinfinity}
G^{\infty}_0&=&\left(\frac{\mu}{2}\right)^{4-D}\int\frac{d^3\mathbf k}{(2\pi)^3}\frac{1}{E-\frac{\mathbf{k}^2}{2M}-\frac{(\mathbf{P}-\mathbf{k})^2}{2M}+i\epsilon}
\end{eqnarray} 
which is linearly divergent. In order to preserve Galilean invariance and maintain a sensible power-counting scheme for NR theories with an unnaturally large scattering length, the power-divergence subtraction (PDS)  scheme is used to regularize the  integral \cite{pds,pds2,pds3}. Using PDS, the integral above becomes
\begin{eqnarray}
\label{IinfinityPDS}
 G^{\infty}_0&=&-\frac{M}{4\pi}\left(\mu+i\sqrt{ME-{\mathbf{P}}^2/4}\right)=-\frac{M}{4\pi}\left(\mu+iq^*\right),
\end{eqnarray}
where $\mu$ is the renormalization scale. When the volume is finite, the integral above is replaced by its FV counterpart, $G^{V}_{0}$. It is straightforward to find the relation between the FV correction $\delta G^{V}_{0}=G^{V}_{0}-G^{\infty}_{0}$ and the NR version of the kinematic function $c_{00}^{P}$ defined in Eq. (\ref{clm}),
\begin{eqnarray}
c^{{P}}_{00}(q^{*2})=\left[\frac{1}{L^3}\sum_{\textbf{k}}-\mathcal{P}\int\frac{d^3\mathbf{k}}{(2\pi)^3}\right]\frac{1}{({\mathbf{k}-\frac{\mathbf{P}}{2}})^{2}-q^{*2}} \ ,
\end{eqnarray}
where we have used the fact that for degenerate particles $\alpha$ that is defined after Eq. (\ref{clm}) is $\frac{1}{2}$, and in the NR limit the relativistic factor $\gamma$ is equal to one.
One can arrive at the desired relation by adding and subtracting the infinite volume two-particle propagator, Eq. (\ref{Iinfinity}), to $G^{V}_{0}$. One of them can be evaluated using PDS, Eq. (\ref{IinfinityPDS}), and the other one can be written in terms of a regularized principle value integral, leading to 
\begin{eqnarray}
\label{Ifinite}
G^V_0(E,{P})
&=&-\frac{M}{4\pi}\mu
-c_{00}^{P}(q^{*2}),
\end{eqnarray}
therefore arriving at
\begin{eqnarray}
\label{fvloop}
\delta G^V_0(E,{P})&=& 
 =\frac{M}{4\pi}\left(q^*\cot\phi^{P}
 +i q^*\right),
\end{eqnarray}
where we have used the pseudo-phase definition, Eq. (\ref{pseudophase}).

The goal is to find a relation between the FV matrix elements of the axial-vector current and the LECs that parametrize the weak interaction, namely $\{g_A,{L_{1,A}}\}$, following a procedure analogous to section~\ref{relLL}. The first step is to find the QC satisfied by the energy eigenvalues of the two-nucleon system in  presence of an external weak field as was also considered in Ref.~\cite{nnd}.\footnote{The main distinction between the result that will be obtained here and that of Ref. \cite{nnd} is that we will consider the case where the two-nucleon system has arbitrary momentum below inelastic thresholds, while Ref. \cite{nnd} only considered the two-nucleon system at rest.} After obtaining the QC for this theory, the trick by Lellouch and L\"uscher~\cite{LL1} can be utilized to obtain an expression for the FV weak matrix element. The main difference between the problem considered here and the problem discussed in the previous section is that the dominant contribution to the weak processes in the NN sector comes from the one-body current, namely the term proportional to the axial charge in Lagrangian, Eq.~(\ref{pionlag}). In fact this contribution modifies the nucleon propagator and therefore the on-shell condition. To avoid complications associated with the modification of \emph{external legs} appearing in the FV analogue of the scattering amplitude, $\mathcal{M}^{V}$, we obtain the QC for this system by looking at the pole structure of the NN correlation function in the presence of the weak field. As before a $2\times2$ kernel $\mathcal{K}$ can be formed that incorporates the tree-level $2\rightarrow2$ transitions, 
\begin{eqnarray}
i\mathcal{K}&=&\begin{pmatrix} 
-i\sum C^{\left(\triplet\right)}_{2n}q^{*2n} &-i{L_{1,A}}\\
-i{L_{1,A}}&-i\sum  C^{\left(\singlet\right)}_{2n}q^{*2n}\\
\end{pmatrix}.
\end{eqnarray}
The FV function $\mathcal{G}^V$ can be still expressed as a $2\times2$  matrix in the basis of channels, except it will attain off-diagonal elements due to the presence of the single-body operator in contrast with the the scalar sector studied before, 
\begin{eqnarray}
\delta\mathcal{G}^V=\left(\begin{array}{ccc}
G_{+}^{V} &  & G_{-}^{V}\\
\\
G_{-}^{V} &  & G_{+}^{V}
\end{array}\right),
\end{eqnarray}
where FV functions $G^V_+$ and $G^V_-$ are defined as
\begin{eqnarray}
\label{GVpm}
G^{V}_{\pm}=\frac{M}{2L^3}\sum_{\mathbf k}\left[\frac{1}{E-\frac{\mathbf{k}^2}{2M}-\frac{(\mathbf{P}-\mathbf{k})^2}{2M}-W_3g_A}\pm
\frac{1}{E-\frac{\mathbf{k}^2}{2M}-\frac{(\mathbf{P}-\mathbf{k})^2}{2M}+W_3g_A}
\right].
\end{eqnarray}
Since we only aim to present the result up to NLO in the EFT expansion according to the power counting discussed above, it suffices to keep only the LO terms in $g_A$ when expanding these FV functions in powers of the weak coupling. Explicitly, $G^V_+=G^V_0(E,{P})+\mathcal{O}(W_3^2g_A^2)$ where $G^V_0$ is defined in Eq.~(\ref{Ifinite}), and $G^V_-=W_3g_A~G^V_1(E,{P})+\mathcal{O}(W_3^3g_A^3)$ with
\begin{eqnarray}
\label{GV1}
G^{V}_1=\frac{M}{L^3}\sum_{\mathbf k}\frac{1}{((\mathbf{k}-\frac{\mathbf{p}}{2})^2-q^{*2})^2}.
\end{eqnarray}

\begin{figure}
\begin{center}
\includegraphics[totalheight=5.0cm]{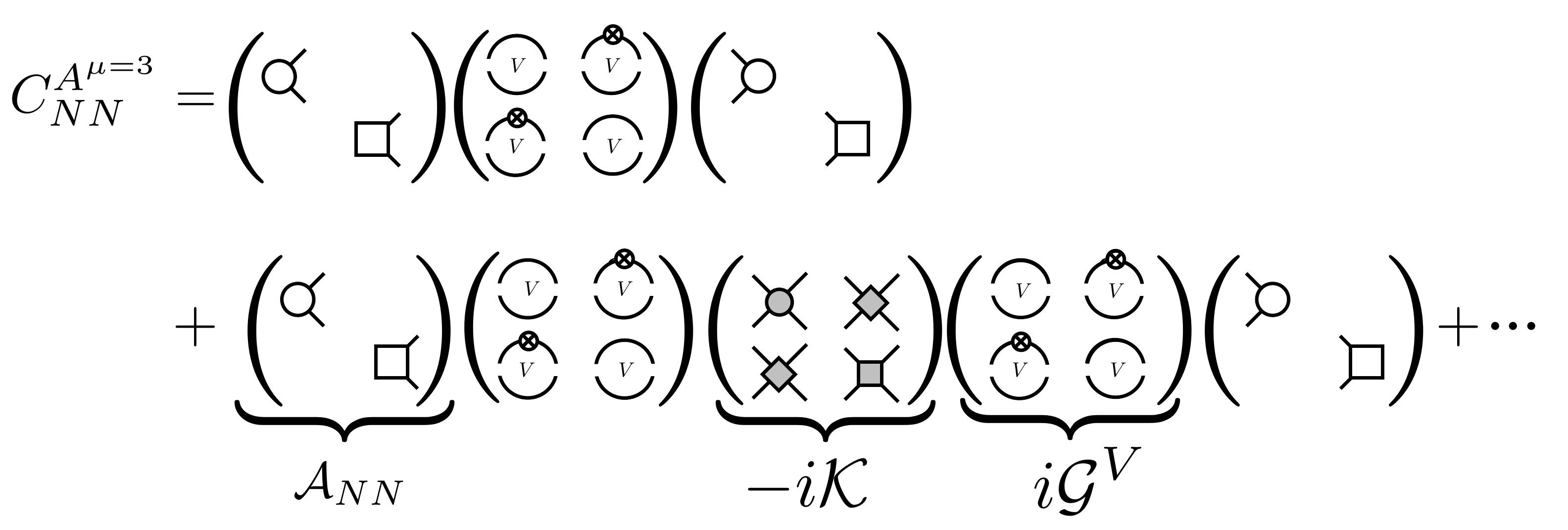}
\end{center}
\caption{ Shown is the NN correlation function in the isosinglet (isotriplet) channel in the presence of an external weak field. $\mathcal{A}_{NN}$ denotes the overlap between the NN interpolating operators and the vacuum. The two-dimensional kernel is denoted by $\mathcal{K}$. The diagonal terms of the kernel correspond to the strong part of the interactions, while off-diagonal terms depict contributions that arise from the weak interaction, namely $L_{1,A}$. Unlike the scalar sector considered before, the FV function, $\delta\mathcal{G}^V$, has diagonal and off-diagonal contributions due to the presence of the single-body current. }\label{CC_ppfusion}
\par\end{figure}

In order to form the NN correlation function, let us also introduce a diagonal matrix $\mathcal{A}_{NN}$, whose each diagonal element denotes the overlap between the two-nucleon interpolating operators in either isosinglet or isotriplet channels and the vacuum. With theses ingredients, the NN correlation function in the presence of the external weak field can be easily evaluated, as is diagrammatically presented in Fig.~(\ref{CC_ppfusion}). It is important to note that in evaluating the FV loops, one should pay close attention to the pole structure of $G^V_{\pm}$, Eq.~(\ref{GVpm}). In other words, the on-shell condition for the free two-nucleon system is modified in the presence of the single-body weak current, namely, $q^{*2}\rightarrow q^{*2}\pm MW_3g_A$.
Then it is straightforward to show that after keeping only terms up to $\mathcal{O}(C_2q^{*2}W_3g_A,W_3L_{1,A})$, the QC obtained from the pole structure of the NN correlation function reads
\begin{eqnarray}
\label{1bLONRquant}
\left[q^*\cot\delta^{(\singlet)}+q^*\cot \phi^P\right]
\left[q^*\cot\delta^{(\triplet)}+q^*\cot \phi^P\right] 
=\left[\frac{4\pi}{M}W_3{\widetilde{L}_{1,A}}
+\frac{4\pi}{M}W_3g_AG_1^V\right]^2,
\end{eqnarray}
where ${\widetilde{L}_{1,A}}$ that is defined as 
\begin{eqnarray}
{\widetilde{L}_{1,A}}&=&\frac{1}{C^{\left(\triplet\right)}_0C^{\left(\singlet\right)}_0}\left[{{L}_{1,A}}-\frac{g_AM}{2}\left(C^{\left(\triplet\right)}_2+C^{\left(\singlet\right)}_2\right)
\right],
\end{eqnarray}
is a renormalization scale independent quantity~\cite{Butler:1999sv, Butler:2000zp, Butler:2001jj,nnd}
\begin{eqnarray}
\mu \frac{d}{d\mu}{\widetilde{L}_{1,A}}=0.
\end{eqnarray}

Before proceeding let us compare this result with the one presented in Ref.~\cite{nnd}. As discussed, the authors of Ref. \cite{nnd} have obtained the same quantization condition for two-nucleon systems in the presence of an external weak field using a dibaryon formalism. The advantage of this formalism is that the diagrammatic representation of the processes of interest are greatly simplified using an auxiliary field with quantum numbers of two nucleons. In fact, the full di-nucleon propagator sums up all $2\rightarrow2$ interactions nonperturbatively.
In the QC presented in Ref. \cite{nnd}, the contributions of the axial charge current to all orders have been kept, but as the higher order operators that contribute to the weak transition have not been included in their calculation, their result is only valid up to $\mathcal{O}(g_AC_2q^{*2})$~\cite{Butler:1999sv, Butler:2000zp, Butler:2001jj}. In the dibaryon formalism, the two-body weak current is parametrized by $l_{1,A}$ which is related to ${\widetilde{L}_{1,A}}$ in this work via\footnote{Note that Eq.~(31) of Ref.~\cite{nnd}  defines $l_{1,A}$ as $\frac{8\pi}{M}{\widetilde{L}_{1,A}}$, but we suspect this discrepancy is only due to a typo in their result. 
}
\begin{eqnarray}
l_{1,A}=\frac{16\pi}{M}{\widetilde{L}_{1,A}}.
\end{eqnarray}
Using this relation between the LECs of both theories, and keeping in mind the order up to which the resuslt of both calculations are valid, one will find agreement between the result presented in Eq. (\ref{1bLONRquant}) and that of Ref.~\cite{nnd} after setting the momentum of the CM to zero.

Having obtained the QC for this system, Eq.~(\ref{1bLONRquant}), it is straightforward to obtain the relationship between the FV matrix elements of the Hamiltonian density and the LECs. In the absence of weak interactions, the two NN states are degenerate with energy $E_0^*$ and on-shell momentum $q^*_0$, satisfying the free quantization condition $\cot(\phi^P)=-\cot(\delta)$. As the weak interaction is turned on, the degeneracy is lifted, leading to a shift in energy equal to $\Delta E^*=V|\mathcal{M}^V_{\singlet-\triplet}|$, where $|\mathcal{M}^V_{\singlet-\triplet}|$ is the FV matrix element of the Hamiltonian density between the $^1S_0$ and $^3S_1$ states. Note that this matrix element is proportional to $W_3$. Therefore it is convenient to define the purely hadronic matrix element $|\mathcal{M}^V_{W}|=|\mathcal{M}^V_{\singlet-\triplet}|/W_3$ which is in fact what would be calculated via LQCD. Expanding the Eq. (\ref{1bLONRquant}) about the free energy, and keeping LO terms in the weak interaction, one obtains
\begin{eqnarray}
\label{1bNRLL}
\left(\frac{MV}{2}\right)^2\csc^2{\delta^{(\singlet)}}\csc^2{\delta^{(\triplet)}}
\left(\phi'+\delta^{(\singlet)'} \right)\left(\phi'+\delta^{(\triplet)'}\right)|\mathcal{M}^V_{W}|^2
=\left(\frac{4\pi}{M}
{\widetilde{L}_{1,A}}+\frac{4\pi}{M}g_AG_1^V\right)^2.
\end{eqnarray}
This result shows that in order to determine weak matrix elements in the NN sector,  not only it is necessary to determine the derivatives of the phase shifts in the $\singlet$ and $\triplet$ channels with respect to the on-shell momenta, but also it is necessary to determine the nucleon axial coupling constant. There is no clear crosscheck for this result, since it is not clear how to implement the density of states approach for this problem. The presence of the one-body operator makes the mixing between the two states non-trivial, therefore one would expect a more complicated relationship between the FV and  infinite volume states than the one predicted via the density of states approach.  Although it would be desirable to obtain a generalization of Lellouch and L\"uscher's result for $2\rightarrow 2$ systems, this example demonstrates that in the two-body sector, one-body currents lead to large FV corrections. In principle, the FV matrix elements depend on the nature of the problem that is considered, and each weak hadronic process must be separately studied.

 \section{Summary and Conclusion}
In this paper, we have presented and derived the FV quantization condition for a system of multi-coupled channels each composed of two hadrons in a moving frame. In the second section, the quantization condition at LO in mixing phase, Eq. (\ref{LOquant}) is derived for the S-wave scattering, where a rather simple EFT toy model for scalar particles is used. Using the techniques developed by Kim $et\; al.$ \cite{sharpe1}, the quantization condition for the boosted systems with arbitrary mixing angles, Eq. (\ref{det0}) has been obtained. It then became evident that the toy model used is in perfect agreement with the nonperturbative result when the angular momenta is truncated at the S-wave. From the generalized results, we have also derived the quantization condition for N=3 coupled channels composed of two hadrons, Eq. (\ref{det3}). The advantage of this result is that it allows the lattice practitioners to perform coupled-channel calculations at multiple boosted momenta in a periodic volume, thereby increasing the number of measurements in order to best constrain the S-matrix elements. When N=2, the S-matrix can be parametrized by three real parameters (two scattering phase shifts and one mixing angle), therefore one needs to perform at least three measurements at each CM momentum, which can be done by using combinations of different boost momenta and different volumes. The N=3 case would require six measurements for each CM momentum to constrain the three phase shifts and three mixing angles.

We have also derived the relationship between FV matrix elements and infinite volume matrix elements in the two-body sector, Eq. (\ref{mesonLL}). This was first calculated for $K\rightarrow \pi\pi$ in the rest frame by Lellouch and L\"uscher  \cite{LL1}, and later extended to the boosted system in Refs. \cite{sharpe1, movingframe2}. Here we have shown two ways to obtain the extension of LL-factor for $2\rightarrow 2$ relativistic processes. The first entails expanding the coupled quantization,  Eq. (\ref{LOquant}), about the free energy of the system when the two channels are decoupled. This method assumes the two states to be degenerate in the absence of the weak interaction, as was first developed by Lellouch and L\"uscher \cite{LL1}. Kim $et\; al.$ \cite{sharpe1} generalized the method of  \cite{LL2} and derived the relationship between FV and infinite volume two-particle states in the moving frame, Eq. (\ref{FVstates}), which has been shown to agree with our result of the two-body relativistic LL-factor, Eq. (\ref{mesonLL}). This derivation strongly depends on the fact that the two channels only mix in the presence of two-body currents. In reality most systems are in fact sensitive to one-body currents as well.  

We have used EFT($\pinot$)  \cite{pionless,pds, pds2, pionless2, pionless3, pionless3} to determine the extension of the LL-factor for NR baryonic systems. In particular, we consider processes that mix the $\singlet-\triplet$ NN channels. This is pertinent for performing calculations of proton-proton fusion, among other interesting processes, directly from LQCD,  \cite{nnd}. The channels in this system are mixed not only by a two-body operator but also by a one-body operator. As it is shown, FV effects arising from the insertion of a one-body operator are sizable and therefore must be included. Unlike any previous case, the FV and infinite volume weak matrix elements are not simply proportional to each other. The result demonstrates that in fact the FV matrix element is proportional to a linear combination of the LO and NLO LECs that parametrize the weak interactions in the NN sector, which can in turn be used to constrain the proton-proton fusion transition amplitude at the few-percent level~\cite{Kong:1999tw, Kong:1999mp, Butler:2000zp}.

\subsection*{Acknowledgment}

We would like to thank Martin J. Savage and William Detmold for fruitful discussions, and for the feedback on the first manuscript of this paper. We also thank Stephen R. Sharpe and Maxwell T. Hansen for useful conversations, and for communicating their results with us prior to publication. RB and ZD were supported
in part by the DOE grant DE-FG02-97ER41014.

\bibliography{bibi}
\end{document}